\documentclass[aps,twocolumn,prl,showpacs,preprintnumbers,amsmath,amssymb]{revtex4}
\def\etal{{\it et\ al.}}

\newcommand{\lsim} 
 {\ \raise.35ex\hbox{$<$}\kern-0.75em\lower.5ex\hbox{$\sim$}\ }
\newcommand{\gsim}
 {\ \raise.35ex\hbox{$>$}\kern-0.75em\lower.5ex\hbox{$\sim$}\ }
\def\journal #1#2#3#4{#1 {\bf #2}, #3 (#4)}

\def\PRB{Phys.\ Rev.\ B}
\def\PRL{Phys.\ Rev.\ Lett.}

\def\SSC{Solid State Commun.}

\def\JPCS{J.\ Phys.\ Chem.\ Solids}

\def\JPSJ{J.\ Phys.\ Soc.\ Jpn.}

\def\SSC{Sol.\ St.\ Comm.}

\usepackage{graphicx}
\usepackage{dcolumn}
\usepackage{bm}
\usepackage{amsmath}
\usepackage{times}
\usepackage[usenames]{color}

\begin{document}

\title{Quantum fluctuation and geometrical frustration effects on electric polarization}
\author{ Tsutomu~Watanabe$^{1 \ast}$, and Sumio~Ishihara$^{2}$ }
\affiliation{$^1$Institute of Multidisciplinary Research for Advanced Materials, Tohoku University, Sendai 980-8577, Japan \\
$^2$Department of Physics, Tohoku University, Sendai 980-8578, Japan 
}
\date{\today}
\begin{abstract}
We examine theoretically a possibility of ferroelectricity caused by electronic charge order without inversion symmetry, motivated by layered iron oxides. Quantum electronic models in a paired-triangular lattice are analyzed by utilizing the variational Monte Carlo simulation. 
Our calculation demonstrates that combined effects of electron transfer between the layers, corresponding to quantum fluctuation between the potential minima, and geometrical frustration promote appearance of an electric polarization. 
Present results are in contrast to the conventional manner of quantum fluctuation 
in ferroelectricity. 
\end{abstract}


\maketitle



%
%

%




In solid state textbook, it has been explained that ferroelectric transition in the displacive-type 
ferroelectricity is caused by a balance of the 
long-range Coulomb interaction and the 
short-range repulsion between ions. 
In addition to these classical explanation, 
recently developed ab-initio band calculations reveal that 
the electron covalency between cations and anions have considerable  
contributions in the ferroelectric transition. 
As for the order-disorder type ferroelectricity, 
the classical dipole-dipole interaction aligns the permanent dipoles and brings 
about the ferroelectric transition. 
Quantum effect plays important roles on the hydrogen-bond 
type ferroelectricity \cite{Blinc,deGennes}. 
The quantum proton tunneling between the two potential minima suppresses the transition, 
and arouses the quantum phase transition between the ferroelectric 
and quantum paraelectric phases. 

\begin{figure}[t]
\vspace{-0.2cm}
\begin{center}
\includegraphics[width=8.0cm,height=1.2cm]{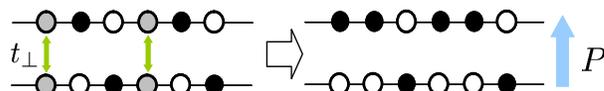}
\end{center}
\vspace{-0.6cm}
\caption{(Color online) 
A three-fold charge order structure and the electric polarization. 
Black, white and gray circles represent charge rich, poor and neutral sites, respectively. 
}
\label{fig:fig1}
\vspace{-0.6cm}
\end{figure}

There is another class of ferroelectricity where an electric dipole and a ferroelectric transition 
are attributed to the electronic origin. 
This is termed an electronic ferroelectricty. 
For this type of the ferroelectricity, 
several experiments in transition-metal oxides \cite{Kato,Ikeda,Tokunaga} 
and organic salts \cite{Monceau,Yamamoto,horiuchi} have been reported, 
and some exotic theoretical ideas have been proposed~\cite{Efremov,Brink}. 
Most of the systems belong to the charge-ordered insulator with the so-called 1/4 
filling, and a charge alignment without the inversion symmetry is responsible for the electric 
dipole moments. 
Here we give a simple example in Fig.~\ref{fig:fig1}; a charge order (CO) model in a pair of chains or layers. 
In the CO with the three-fold or odd-fold periodicity, 
a pair of the charge-neutral sites, 
represented by the grey circles, is sandwiched between the checker-board type 
charge configuration. 
Quantum charge fluctuations are caused by the electron transfer between the two potential minima. When the electrons are localized in one side due to some cooperative interactions, the macroscopic electric dipole appears. 

Instead of extensive material syntheses and experimental examinations in the electronic ferroelectricity,
the microscopic mechanisms have not been clarified yet. 
The long-range Coulomb interaction may promote    
an electronic charge ordering associated with electric polarization. 
In comparison with other types of ferroelectricity, 
quantum fluctuation due to electron motion should be crucially important, because the electron mass is much smaller than  
the proton and ion masses. 
Large fluctuation causes the diffusive nature in the dielectric constant and the optical spectra observed commonly in this class of material~\cite{Ikeda,Monceau,Xu}. 
It is expected naively that such fluctuation prevents a system from the long-range dielectric order, 
on the analogy of the quantum phase transitions in the hydrogen-bond type ferroelectricity and SrTiO$_3$. 

In this Letter, we demonstrate, against such conventional expectation, that, in a frustrated geometry, a quantum 
charge fluctuation promotes the electric polarization. 
The present study is motivated by an electronic ferroelectric material, the layered iron oxide $R$Fe$_2$O$_4$ ($R$=Lu, Yb, Y). 
The crystal structure is made up of a stacking of the paired-triangle lattices~\cite{Kimizuka}, 
as shown in Fig.~\ref{fig:fig4}(a). 
The nominal valence of a Fe ion is 2.5+ implying a coexisting of an equal amount of Fe$^{2+}$ and Fe$^{3+}$. 
The electric polarization measured by the pyroelectric current is enhanced around the charge ordering temperature of the three-fold \cite{Ikeda,Yamada,Mulders}. 
A charge disproportionation in the paired-triangle layers is believed to be responsible for the electric dipole moment, 
as expected in Fig.~\ref{fig:fig1}. 
Here we concentrate on a paired-triangle lattice, 
and take up a possibility of the electronic ferroelectricity. 
In particular, as well as the quantum charge fluctuation, 
we focus on roles of the geometrical frustration which is often seen in the electronic ferroelectric materials~\cite{Kato,Ikeda,Yamamoto}.  
To explore this issue, the three electronic models are analyzed by the numerical method of the variational Monte-Carlo (VMC) simulation. 
We show that the electron transfer between the layers, corresponding to the quantum fluctuation between the potential minima, stabilizes the electric polarization, when the system is strongly frustrated.

\begin{figure}[!t]
\vspace{-0.2cm}
\begin{center}
\includegraphics[width=1.02\columnwidth,clip]{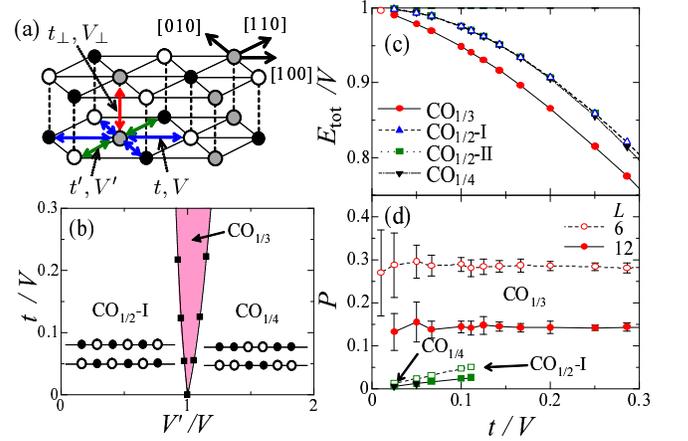}
\end{center}
\vspace{-1.5cm}
\caption{(Color online) 
(a) A paired-triangular lattice adopted in the Hamiltonian ${\cal H}_{Vt}$.
(b) Phase diagram for ${\cal H}_{Vt}$.  
Relations $t=t'=t_\bot$ and $V=V_\bot$ are assumed. 
(c) The energies for the several CO patterns at $V'/V=1$. 
(d) The polarization correlation $P$. 
Circles, squares and inverse triangles are for the results in 
CO$_{1/3}$ at $V'/V=1$, CO$_{1/2}$ at $V'/V=0.95$ 
and CO$_{1/4}$ at $V'/V=1.05$, respectively. 
Open and filled symbols are for the results in L=6 and 12, respectively. 
}
\label{fig:fig2}
\vspace{-0.5cm}
\end{figure}

We start from a possibility of the electronic ferroelectricity 
in a simple spin-less fermion model where the long-range Coulomb interactions and the electron transfers are taken into account.  
We consider a pair of the triangular layers where 
two triangular lattices are regularly stacked along the $c$ axis as shown in Fig.~\ref{fig:fig2}(a); 
a site in the upper layer is just upon a site in the lower one. 
Effects of the realistic crystal structure of $R$Fe$_2$O$_4$ and spin degree of freedom  
will be considered later.  
%
The Hamiltonian is given as  
${\cal H}_{Vt} = {\cal H}_t + {\cal H}_V$ 
with 
\begin{eqnarray}
{\cal H}_t= 
&-&t  \sum\nolimits_{<ij>  m} c_{m i}^{\dag} c_{m j} 
-t' \sum\nolimits_{(ij)  m} {c_{m i}^{\dag} c_{m j} } \nonumber\\ 
&-&t_\bot \sum\nolimits_{i } c_{u i}^{\dag } c_{l i} +H.c. , 
\label{eq:Ht} 
\end{eqnarray}
and 
\begin{eqnarray}
{\cal H}_V &=& 
 V  \sum\nolimits_{<ij> m} n_{m i} n_{m j}  
+V' \sum\nolimits_{(ij) m} n_{m i} n_{m j} \nonumber \\ 
&+&V_\bot \sum\nolimits_i n_{u i} n_{l i}, 
\label{eq:Hv} 
\end{eqnarray}
where 
$c^\dag_{m i}$ is the creation operator for 
a spin-less fermion at site $i$ on the upper layer $(m=u)$ 
or the lower one $(m=l)$, 
and $n_{m i}= c_{m i}^\dag c_{m i}$ is the number operator. 
A subscript $i$ takes a two-dimensional coordinate in a triangle lattice. 
To examine roles of frustration, 
we consider an anisotropic triangular lattice where bonds along the [110] direction are inequivalent to 
those along [100] and [010] [see Fig.~\ref{fig:fig2}(a)]. 
Introduced $V$, $V'$ and $V_\bot$ are the Coulomb interactions between the nearest-neighboring (NN) sites along the [100] and [010] directions, the [110] one, and the [001] one, respectively. 
Symbols $\sum_{\left\langle {ij} \right\rangle}$ and 
$\sum_{\left( {ij} \right)}$ 
imply summations for the NN pairs.  
The three transfer integrals, $t$, $t'$ and $t_\bot$, 
are defined in the same way. 
Ratios $V'/V$ and $t'/t$ express measure of frustration. 

Our optimization VMC method can perform 
unbiased calculation where both the frustration and quantum effects 
are treated accurately \cite{Umrigar}. 
We adopt 2-5$\times$10$^5$ samples in most of the simulations where  
an error of the energy expectation value is of the order of $10^{-4}t$. 
Systems used are of $N_s = L\times L\times 2 (\equiv 2N)$ $(L\le 12)$ 
sites with the periodic- and antiperiodic-boundary conditions, 
and the particle number is $N_e=N_s/2$. 
A variational wave function is defined as $\Psi={\cal P}\Phi$, 
with the one-body Hartree-Fock (HF) part $\Phi$ and 
the many-body correlation factor ${\cal P}$. 
We define 
${\cal P}\equiv \prod\nolimits_{ij} {\left ({1-v_{ij} n_i n_j} \right)}$ 
where $v_{ij}$ takes one of the three variational parameters,    
$v$, $v'$ and $v_\bot$, 
when the NN $ij$ pairs are along the [100] and [010] directions, 
the [110] one and the [001] one, respectively. 
As for $\Phi$, we assume the HF wave functions for the 
following four-types of the CO structures; 
(i) a two-fold CO along the [110] direction termed CO$_{1/2}$-I, 
(ii) a two-fold CO along [100] termed CO$_{1/2}$-II, 
(iii) a four-fold CO along [110] termed CO$_{1/4}$, and 
(iv) a three-fold CO along [110] termed CO$_{1/3}$ 
[see Fig.~\ref{fig:fig2}(a) and the inset of Fig.~\ref{fig:fig2}(b)]. 
In each CO, 
a magnitude of the order parameter is introduced as a variational parameter $\Delta$ in $\Phi$. 
In the CO$_{1/2}$-I, CO$_{1/2}$-II and CO$_{1/4}$ structures, stackings of the CO patterns in the two triangular-layers are set to be out-of-phase. 
In CO$_{1/3}$, the assumed pattern along the $[110]$ direction is schematically 
$\cdots \bullet \circ \star \bullet \circ \star \cdots$ 
($\cdots \circ \bullet \star \circ \bullet \star \cdots$) 
in the upper (lower) plane.  
Symbols $\bullet$, $\circ$ and $\star$ imply the charge densities of  
$\langle n_{m i} -1/2 \rangle=\Delta$, $-\Delta$ and zero, respectively. 

In Fig.~\ref{fig:fig2}(a), we present the CO phase diagram. 
We set up $t=t'=t_\bot$ and $V=V_\bot$, and 
focus on roles of the frustration, 
$V'/V$, and the electron transfer. 
In the classical limit, i.e. $t=0$, 
the CO$_{1/2}$-I (CO$_{1/4}$) structure are realized 
in the regions of $V'/V<1$ ($>1$). 
The CO$_{1/2}$-II structure is degenerate with CO$_{1/4}$. 
The CO$_{1/3}$ structure of the present interest 
appears only at $V'/V=1$ where 
all CO's are degenerate. 
By introducing $t$, 
CO$_{1/3}$ appears in a finite parameter region. 
That is, CO$_{1/3}$ is stabilized by a combined effect of the 
quantum fluctuation and the geometrical frustration. 

The energy expectation 
$E_{\rm tot} \equiv N^{-1} \langle {\cal H}_{Vt} \rangle$ is presented as 
a function of $t/V$ at $V'/V=1$ in Fig.~\ref{fig:fig2}(c). 
It is clearly shown that a stabilization of CO$_{1/3}$ is attributed to the $t$-linear dependence of $E_{\rm tot}$, in contrast to the quadratic dependence in other CO's. 
A schematic pattern of CO$_{1/3}$ is shown in Fig.~\ref{fig:fig2}(a). 
The energy gain is caused by the inter-layer electron transfer, $t_\bot$ as follows. 
Let us focus on a pair of the gray circles in the upper and lower layers in Fig.~\ref{fig:fig2}(a). 
This system corresponds to the symmetric double-wells aligned on a triangle lattice. 
Since these sites are surrounded by the in-plane three NN charge-rich and three poor sites, 
the Coulomb interactions are canceled out. 
Thus, the classical energy does not depend on the charge configurations 
at the sites represented by the gray circles in Fig.~\ref{fig:fig2}(a). 
This degeneracy of the order of $2^{N/3}$ is lifted independently at each site by the first order of $t_{\bot}$. 
Even when the longer-range Coulomb interactions than $V$, $V'$ and $V_\bot$ are considered, this scenario survives in the case where 
$t_\bot$ is larger than those interactions.

The electric-polarization correlation defined by 
$P = \left\langle {p^2} \right\rangle^{1/2}$ with 
$p =( 3/N) 
\sum_i
\left(  n_{ui} -  n_{l i}  \right)$ 
is presented in Fig.~\ref{fig:fig2}(d). 
The possible maximum value of $P$ in CO$_{1/3}$ is one. 
The obtained value in CO$_{1/3}$ is much higher than those in other 
CO's. 
However, with increasing the system size $L$, a magnitude of $P$ in 
CO$_{1/3}$ decreases considerably.
We suppose that amplitude of $P$ in the thermodynamic limit is 
much smaller than its maximum value. 


The above results imply that the quantum charge fluctuation and the geometrical frustration promote 
the three-fold CO upon which the electric polarization has a chance to appear.  
However, in the large system size, 
the amplitude of $P$ in the polar state is rather small. 
When we focus on one of the paired-grey sites in Fig.~\ref{fig:fig2}(a), corresponding 
to a symmetric double-well potential, 
the bonding state is stabilized by $O(t_\bot)$, 
and a symmetric charge distribution is realized. 
The present numerical results imply that, 
within this model, 
the effective interactions between the pairs are not enough to induce an asymmetric charge distribution. 
We expect that additional small factors, which are not taken into account so far, may stabilize the polar CO structure. 
We propose two plausible candidates, the spin degree of freedom 
and the realistic crystal structure in $R$Fe$_2$O$_4$. 

We are modeling the spin degree of freedom on the basis of LuFe$_2$O$_4$ where 
two kind spins, $S=2$ $(d^6)$ in Fe$^{2+}$ and $S=5/2$ $(d^5)$ in Fe$^{3+}$, coexist. 
An excess electron in a Fe$^{2+}$ is itinerant and is responsible for CO, 
and others are regarded as a localized spin with $S=2$ because of their  small transfer integral~\cite{Nagano,Naka}. 
The model is made up of the itinerant electrons coupled ferromagnetically with the localized spins 
through the on-site Hund coupling. 
This is a kind of the double-exchange model. 
We assume, for simplicity, the infinite-limit of the Hund coupling and the  
Ising-type localized spins denoted by $\sigma_{m i}$ $(m= u, l)$ which takes $\pm 1$. 
The Hamiltonian is  
$ {\cal H}_S={\cal H}_{t}'+{\cal H}_V+{\cal H}_{\rm AFM}$ 
where ${\cal H}_V$ 
is given in Eq.~(\ref{eq:Hv}).  
The first term is the same form with Eq.~(\ref{eq:Ht}),  
but the all transfer integrals are replaced by the spin dependent forms. 
For example, 
$-t \sum\nolimits_{<ij>  m} c_{m i}^{\dag} c_{m j} 
\rightarrow 
- \sum\nolimits_{<ij>  m} t_{\sigma_{m i} \sigma_{m j}} c_{m i}^{\dag} c_{m j} $
where $t_{\sigma_{mi} \sigma_{mj}}=t$ for $\sigma_{mi}=\sigma_{mj}$ and 
zero for $\sigma_{mi} \ne \sigma_{mj}$. 
The last term ${\cal H}_{\rm AFM}$ is the antiferromagnetic interaction between the localized spins,  
${\cal H}_{\rm AFM}=J_{S} \{\sum\nolimits_{m <ij>}  \sigma_{mi} \sigma_{mj} 
 + \sum\nolimits_{(ij) m} \sigma_{mi} \sigma_{mj} + \sum\nolimits_i {\sigma_{ui} \sigma_{li}} \}$ 
with the positive exchange constant $J_S$. 
We analyze this model by using the VMC method. 
We consider all possible magnetic orders in the localized-spin sector 
characterized by the momenta $(2\pi/3, 2\pi/3, 0)$ which corresponds to the observed neutron diffraction peak at $(1/3\ 1/3\ m)$ in LuFe$_2$O$_4$ \cite{Akimitsu,Shiratori,Angst,Wu}. 
We restrict our calculations to the most frustrated case, i.e. $V'/V=1$ and 
$t'/t=1$. 

\begin{figure}[t]
\vspace{-0.2cm}
\begin{center}
\includegraphics[width=0.7\columnwidth,clip]{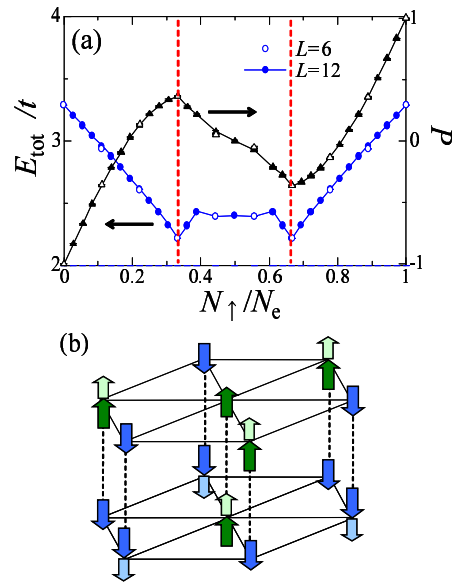}
\end{center}
\vspace{-0.7cm}
\caption{(Color online)
(a) The energy and the electric polarization correlation in 
the model with the spin degree of freedom, ${\cal H}_S$.  
Relations $t=t'=t_\bot$ and $V=V'=V_\bot$ are assumed. 
Parameter values are chosen to be $t/V=0.2$ and $J_S/t=0.05$. 
Open and filled symbols are for the results in $L=6$ and 12, respectively. 
(b) A schematic view of the optimized spin structure.  
Short and long arrows represent spin directions of the conduction electrons and the localized spins, 
respectively. 
}
\label{fig:fig3}
\vspace{-0.5cm}
\end{figure}

In Fig.~\ref{fig:fig3}(a),  
the energy and the polarization correlation in the optimized spin state are plotted as a function of $N_\uparrow/N_e$. 
Here, $N_{\uparrow (\downarrow)}$ is the number of the conduction electrons 
with spin up (down) and $N_e=N_\uparrow+N_\downarrow$. 
The energy minima are seen at $N_\uparrow/N_e=0.33$ and $0.67$ 
($N_\downarrow/N_e=0.33$), where a large value of $P$ ($\sim 0.37$) 
is realized. 
A magnitude of $P$ is 
almost independence of the system size. 
That is, the polar state becomes robust by the spin ordering. 

This stabilization of the polar state 
is attributed to the so-called gate effect due to spin degree of freedom as follows. 
The optimized spin structure is presented in Fig.~\ref{fig:fig3}(b). 
It is easily seen that, without the spin degree of freedom, 
the classical energy is not changed by an electron motion from a site in the lower layer to the one just upon it. 
The energy gain in this process is proportional to $t_{\bot}$, being  similar to the previous situation in the CO$_{1/3}$. 
In contrast to the previous case, the spin order suppresses the electron hopping, and keeps the electrons in the lower layer. 
This effect is seen, for example, for the down-spin electrons in the lower layer; since the number of the down localized-spins in the lower layer is 
twice of that in the upper one, the down-spin electrons are confined in the lower layer to gain the in-plane kinetic energy. 
In other word, the itinerant electrons are magnetically polarized by the localized spins. As a result, charge imbalance is induced between the layers, and the electric polarization appears. 
It is worth noting that the spontaneous spin- and charge-polarizations in the conduction electrons 
are not caused by an asymmetric potential due to the localized spins; 
the optimized localized-spin configuration is invariant under the space and time inversions plus the in-plane translation along [100].
 

Another candidate to realize the polar CO state is the 
realistic crystal structure in $R$Fe$_2$O$_4$. 
Unlike in the models considered so far, in $R$Fe$_2$O$_4$, 
a position of a Fe ion in the upper layer is not just upon 
a Fe ion in the lower layer, but is upon an O ion surrounded by three in-plane Fe ions [see Fig.~\ref{fig:fig4}(a)]. 
We modify the previous Hamiltonian ${\cal H}_{Vt}$ by 
considering this crystal structure. 
We introduce the long-range Coulomb interactions for the in-plane NN sites 
($V_{\rm abNN}$), that for the inter-plane NN ones ($V_{\rm cNN}$), and 
for the inter-plane NNN ones ($V_{\rm cNNN}$). 
When the $1/r$-type Coulomb interaction is assumed, 
we have $V_{\rm cNN}/V_{\rm abNN}=1.2$ and $V_{\rm cNNN}/V_{\rm abNN}=0.77$ 
\cite{Kimizuka}. 
As for the electron transfer term ${\cal H}_t$, the transfer integral for the in-plane NN sites ($t_{\rm ab}$), and that for the inter-plane NN ones ($t_{\rm c}$) are introduced.  
This model is analyzed by the VMC method. 
Instead of the CO$_{1/3}$ introduced previously, 
we adopt the polar three-fold CO (polar CO$_{1/3}$) and the six-fold one (CO$_{1/6}$) 
which are degenerate in the HF level. 
The CO patterns 
along $[100]$ is schematically 
$\cdots \bullet \circ \bullet \bullet \circ \bullet \cdots$ 
($\cdots \circ \bullet \circ \circ \bullet \circ \cdots$ ) 
in the upper (lower) plane 
for the polar CO$_{1/3}$, and is 
$\cdots \bullet \circ \bullet \bullet \circ \circ \cdots$ 
($\cdots \circ \bullet \circ \circ \bullet \bullet \cdots$ ) 
for CO$_{1/6}$, 
which is regarded as an antiferro-electric state. 
Symbols $\circ$ and $\bullet$ imply the charge densities of  
$\langle n_{m i}-1/2 \rangle=\Delta$ and $-\Delta$, respectively, 
with the variational parameter $\Delta$.  

\begin{figure}[t]
\vspace{-0.2cm}
\begin{center}
\includegraphics[width=9.0cm,height=6.4cm]{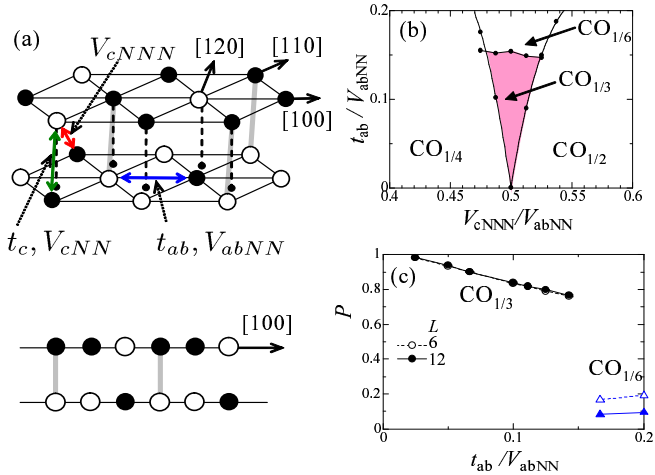}
\end{center}
\vspace{-0.8cm}
\caption{(Color online)
(a) A paired-triangular lattices in $R$Fe$_2$O$_4$ 
and the polar CO$_{1/3}$ structure. 
The lower panel shows a side view along the [120] direction. 
(b) Phase diagram for the model where 
the realistic crystal structure is taken into account. 
Relations $t_{\rm c}/t_{\rm ab}=1$ and $V_{\rm cNN}/V_{\rm abNN}=1$ 
are assumed. 
(c) The electric polarization correlation $P$ at $V_{\rm cNNN}/V_{\rm abNN}=0.5$. 
Open and filled symbols are for the results in $L=6$ and $12$, respectively. 
}
\label{fig:fig4}
\vspace{-0.5cm}
\end{figure}

The phase diagram is shown in Fig.~\ref{fig:fig4}(b) 
in a plane of $V_{\rm cNNN}/V_{\rm abNN}$ and $t_{\rm ab}/V_{\rm abNN}$. 
Relations $V_{\rm cNN}=V_{\rm abNN}$ and $t_{\rm ab}=t_{\rm c}$ are assumed, 
although the qualitative features introduced below are robust around 
these relations.  
In the classical limit of $t_{\rm ab}=0$, 
$V_{\rm cNNN}/V_{\rm abNN}=0.5$ is the fully frustrated point, 
where the polar CO$_{1/3}$ is degenerate with other CO's.  
This is similar to the results in Fig.~\ref{fig:fig2}(b). 
By taking the electron transfer into account, the polar CO$_{1/3}$ 
appears around this point. 
Beyond $t_{\rm ab}/V_{\rm abNN} \sim 0.15$, 
the non-polar CO$_{1/6}$ overcomes 
the polar CO$_{1/3}$. 
The electric polarization presented in Fig.~\ref{fig:fig4}(c) 
shows almost the maximum value in the polar CO$_{1/3}$, 
and its value is independence of the system size $L$. 
On the other hand, a value of $P$ in CO$_{1/6}$ tends to vanish in the thermodynamic limit. 

Key interactions to stabilize the polar CO$_{1/3}$ are the inter-layer charge transfer, $t_{\rm c}$, 
and the long-range Coulomb interaction, $V_{\rm cNNN}$, which is not disregarded in this crystal structure. 
As shown in Fig.~\ref{fig:fig4}(a), $V_{\rm cNNN}$ induces the interaction along the [120] direction between 
the bonds which are represented by the pairs of the black-white circles connected by the grey bars.  
These bonds correspond to the pairs of the grey circles in Fig.~\ref{fig:fig2}(a). 
As a result, asymmetric charge distributions in the bonds overcome the symmetric ones in the bonding states  
which are realized in the previous model of ${\cal H}_{Vt}$. 
Although the dipole moments are induced in the local bonds, 
there is still competition between the polar CO$_{1/3}$ 
and the antiferro-electric CO$_{1/6}$ 
which are in the different configurations of the dipole moments along [100]. 
This degeneracy is lifted by $t_{\bot}$. 
Consider the charge fluctuation at the sites represented by the black (white) circles in the upper (lower) layer in Fig.~\ref{fig:fig4}(a). 
In the inter-layer charge fluctuation by $t_{\bot}$, 
the energy reduction is of the order of $t_c^2/V_{\rm cNNN}$ where the intermediate-state energy $V_{\rm cNNN}$ is the smallest Coulomb interaction in the present model of ${\cal H}_{Vt}$. 
Therefore, this kind fluctuation easily occurs and stabilizes the polar CO$_{1/3}$. 
In other CO's, this fluctuation process is suppressed or prohibited. 

Through the present analyses, 
of the three-type models in a paired-triangular lattice 
we demonstrate that the quantum charge fluctuation due to the electron transfer promotes the electric polarization. 
In particular, the polar phase is stabilized remarkably due to the inter-layer charge fluctuation in the strongly frustrated condition.  
The spin degree of freedom of electron and the realistic crystal structure in $R$Fe$_2$O$_4$ are also the key ingredients for the electric polarization. 
In comparison with the hydrogen-bond type ferroelectricity where the proton tunneling suppresses the long-range order, the charge fluctuation in the present electronic ferroelectricity plays a distinguish role. 
From the present scenario of the ferroelectricity driven by the quantum fluctuation, 
the experimentally observed diffusive dielectric constant~\cite{Ikeda} and the valence fluctuation suggested by the optical and M$\rm \ddot o$ssbauer measurements~\cite{Xu,Nakamura} are naturally understood. 
Here, a classical picture, i.e. the polar charge order stabilized  by the Coulomb interaction, is not applicable, and a fragile electric polarization is of prime essence. 
The present results open a new aspect of the quantum fluctuation in 
ferroelectricity. 

Authors would like to thank S.~Mori, N.~Ikeda H.~Takashima and J.~Nasu for their valuable discussions. This work was supported by JSPS KAKENHI, TOKUTEI from MEXT, and Grand challenges in next-generation integrated nanoscience. 

\noindent
\textcolor{red}{
}

$^{\ast}$ Present address: Chiba Institute of Technology, Tsudanuma, Chiba 275-0016, Japan.

\end{document}